\documentclass[prl,aps,showpacs,twocolumn,superscriptaddress]{revtex4-1}
\usepackage{amsmath,amssymb,graphicx}
\usepackage[final]{hyperref}
\usepackage{amsmath}

\begin{document}

\title{3D super-resolved in vitro multiphoton microscopy by saturation of excitation}

\author{Anh Dung Nguyen}
\email[anhdung@ulb.ac.be]{}
\author{François Duport}
\author{Arno Bouwens}
\affiliation{Universit\'e libre de Bruxelles, Service OPERA Photonique, Avenue F.D. Roosevelt 50 CP194/5, 1050 Brussels, Belgium}
\author{Fr\'ed\'erique Vanholsbeeck}
\affiliation{University of Auckland, The Dodds-Walls Centre for Photonic and Quantum Technologies \& Department of Physics, Private bag 92019, Auckland 1142, New Zealand}
\author{Dominique Egrise}
\author{Gaetan Van Simayes}
\author{Serge Goldman}
\affiliation{Universit\'e libre de Bruxelles, Department of Nuclear Medecine, ULB-Hopital Erasme, 808 Route de Lennik, 1070 Brussels, Belgium}
\author{Philippe Emplit}
\author{Simon-Pierre Gorza}
\affiliation{Universit\'e libre de Bruxelles, Service OPERA Photonique, Avenue F.D. Roosevelt 50 CP194/5, 1050 Brussels, Belgium}


\begin{abstract}
We demonstrate a significant resolution enhancement beyond the conventional limit in multiphoton microscopy (MPM) using saturated excitation of fluorescence. Our technique achieves super-resolved imaging by temporally modulating the excitation laser-intensity and demodulating the higher harmonics from the saturated fluorescence signal. The improvement of the lateral and axial resolutions is measured on a sample of fluorescent microspheres. While the third harmonic already provides an enhanced resolution, we show that a further improvement can be obtained with an appropriate linear combination of the demodulated harmonics. Finally, we present in vitro imaging of fluorescent microspheres incorporated in HeLa cells to show that this technique performs well in biological samples.

\end{abstract}


\maketitle

\section{Introduction}
Since its introduction in 1990 \cite{denk1990two}, multiphoton microscopy (MPM) rapidly evolved to become the workhorse for 3D fluorescence imaging in many laboratories. MPM is arguably one of the most influential recent inventions in optical microscopy because it offers a combination of minimally invasive imaging with high resolution, thin optical sectioning and deep penetration in highly scattering samples such as biological tissues \cite{Pawley2006}.\\
These properties all stem from the process of multiphoton fluorescence excitation. In order to achieve efficient multiphoton excitation, photons need to be concentrated both in space by focussing the excitation beam, and in time by using pulsed laser sources. Fluorescence is therefore only excited in a narrow volume. Moreover, because the excitation spectra of common fluorescent molecules are in the visible spectrum, the wavelengths used for MPM are in the near-infrared such that optical sample damage is much reduced \cite{denk1997photon, Zipfel2003}.

A second important development in optical microscopy has been the development of super-resolution microscopy which achieves resolutions beyond the diffraction limit. In essence, super-resolved imaging requires the sample to respond non-linearly to the illumination light. Such a non-linearity can be achieved by, among other techniques, saturation of excitation \cite{Fujita07} and stimulation emission depletion (STED) \cite{Hell1994}. Bringing super-resolution to MPM has the potential to enable studying biological samples at the depths offered by MPM with a resolution beyond the diffraction limit.

STED has been applied successfully for two-photon microscopy (2PM), but only to improve the lateral resolution \cite{Moneron2009}. Moreover, STED requires the use of a high-intensity visible wavelength depletion beam, increasing the invasiveness of the method. Although STED microscopy can also improve the axial resolution beyond the diffraction limit \cite{Wildanger2009}, it requires a carefully aligned and relatively complex optical system, often including an additional depletion laser source.

In this work, we present a simple approach to achieve 3D super-resolution in MPM, based on saturation of fluorescence excitation \cite{Fujita07}. This method can easily be implemented on an existing multiphoton microscope, as it only requires (a) temporal modulation of the excitation beam using an off-the-shelf intensity modulator, and (b) demodulation of the detected fluorescence signal with an electronic lock-in amplifier. Hence, no additional laser sources or beam shaping optics are required, and only near-infrared light is incident on the sample.

Images of fluorescent microspheres on a coverslip prove that this technique can be implemented successfully in MPM and improves not only the lateral but also the axial resolution. Further experiments with microspheres integrated in HeLa cells produced encouraging results demonstrating the feasibility of this method for \textit{in vitro} studies.


\section{Principle of the method}

The technique proposed in this work is based on the saturation of the probability of excitation of the fluorophore in MPM. Until recently, saturation has been considered harmful for imaging and has been avoided \cite{Cianci04,Visscher94}. It is possible, however, to show that in a saturation regime, the relation between the intensity of excitation and the emitted fluorescence intensity is no longer linear and contains nonlinearities that can be used to improve the lateral as well as the axial resolution \cite{Fujita07,Yamanaka08,Yamanaka13}.

In a two-photon excitation process, assuming a delay between pulses much longer than the fluorescence lifetime, the probability of exciting a fluorophore at a given position with one excitation laser pulse is given by
\begin{equation}
 P_{exc} = 1-e^{-\alpha \sigma_{2PA} P^2 PSF^2 (x,y,z)},
\label{Eq1}
\end{equation}
where $\alpha$ is a coefficient which depends on the pulse's shape and duration, and on the repetition rate of the laser, $\sigma_{2PA}$ is the two-photon absorption cross-section of the fluorophore, $P$ the average power of the excitation beam and $PSF(x,y,z)$ is the illumination point spread function. To extract the nonlinearities that appear when saturation occurs, a temporal modulation of frequency $\nu_m$ is applied to the excitation beam \cite{Fujita07}. The probability of excitation can then be written
\begin{equation}
 P_{exc} = 1- e^{-\alpha \sigma_{2PA}P^2[1+\cos(\omega t)]^2 PSF^2 (x,y,z)},
\label{Eq2}
\end{equation}
where $\omega = 2\pi\nu_m$.

\begin{figure}[h]
\centering
\includegraphics[width=0.4\textwidth]{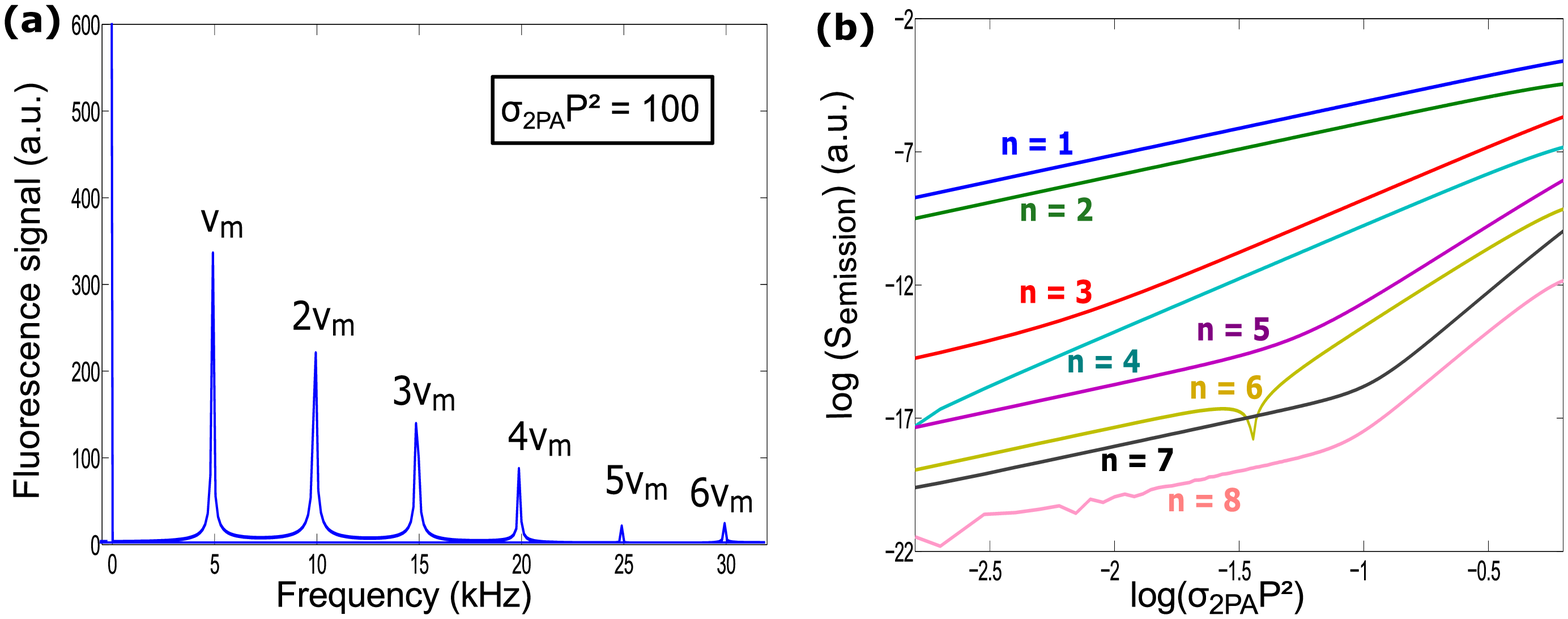}
\caption{Apparition of higher harmonics in the saturation regime. (a) Spectrum of the signal of the detected fluorescence at a modulation frequency of $\nu_m = 5\,$kHz. (b) Power in the $n$-th harmonic of the detected fluorescence signal ($S_{emission}$) for increasing power on the sample.}
\label{Figure1}
\end{figure}

At high excitation power, higher order harmonics in the fluorescence signal will become non-negligible, as illustrated in Fig.~\ref{Figure1}(a).

By developing the $1-e^{-x}$ term up to the $4^{th}$ order in Eq.~(\ref{Eq2}), and rearranging the expression, the power in the harmonics can be calculated
\begin{equation}
\left\{
   \begin{array}{l c r }
     P^{signal}_{\nu_m}   & : & 2 PSF^2 -  \frac{7}{2} PSF^4 + \frac{33}{8} PSF^6 , \\
                     &  &  \\
     P^{signal}_{2\nu_m}  & : & \frac{1}{2} PSF^2  - \frac{7}{4} PSF^4 + \frac{495}{192} PSF^6 ,\\
                    &  &  \\
     P^{signal}_{3\nu_m} & : &  \frac{1}{2} PSF^4 - \frac{55}{48} PSF^6 .\\
   \end{array}
\right.
\label{harmonicsSups}
\end{equation}

These equations show that the harmonics in the detected fluorescence signal contain terms with the fourth and sixth powers of the illumination PSF. In standard two-photon microscopy, the detected signal contains $PSF^2$. Hence, the resolution is improved by at least a factor of $\sqrt{2}$ when using the third harmonic. Figure \ref{Figure1}(b) shows the dependence of the harmonics' strength $S_{emission}$ on the illumination intensity. 

By making the appropriate linear combination of the harmonics in Eq.~(\ref{harmonicsSups}), it is possible to obtain a signal which is dependent only on the sixth and higher powers of the $PSF$:
\begin{equation}
\mbox{Linear combination signal} = P^{signal}_{\nu_m} - 4P^{signal}_{2\nu_m} - 7P^{signal}_{3\nu_m}.
\label{CombiLi}
\end{equation}
The linear combination signal ($\propto PSF^6$) therefore achieves an additional resolution improvement by a factor of $\sqrt{3/2}$ over the third harmonic signal $P^{signal}_{3\nu_m}$ assuming a gaussian illumination profile.


\section{Material and methods}

\subsection{Experimental setup}
\begin{figure}[h]
\centering
\includegraphics[width=0.32\textwidth]{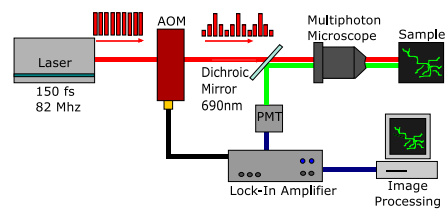}
\caption{The experimental setup is based on a conventional multi-photon microscope with an acousto-optic modulator to modulate the fluorescence excitation intensity, and a lock-in amplifier to extract the harmonics of the fluorescence signal detected by the PMT.}
\label{Figure2}
\end{figure}

The experimental setup, shown in Fig.~\ref{Figure2}, is built around a commercial two-photon microscope (Model FV1000-MPE, based on a BX61WI upright microscope body, Olympus Corp.). The configuration of the microscope is the U-MWG2 which has two detection channels with filters allowing the detection of fluorescence from $510$ to $550\,$nm and from $600$ to $690\,$nm. The microscope objective is the Olympus XLPN25XWMP (numerical aperture $1.05$, magnification $25\times$). The multiphoton excitation laser system used in our experiments is a Spectra-Physics Ti:Sa Tsunami pumped by a Millenia ProS-Series. This laser is tunable over a wavelength range of $840\,$nm to $980\,$nm. It has a repetition rate of $82\,$MHz and is capable of producing $150\,$fs pulses at a wavelength of $880\,$nm. The excitation light beam is temporally modulated by an acousto-optic modulator (MT80-A1-IR, AA Opto Electronic Sa.). A  lock-in amplifier (HF2LI, Zurich Instruments AG) demodulates the signal from the photomultiplier tubes (R3896, Hamamatsu Photonics K.K.) of the microscope at the first, second and third harmonics. The linear combination of harmonic signals is generated by an analog summation circuit such that no offline processing is required to generate the final image. The microspheres used for the different experiments are Dragon Green fluorescent microspheres (Bangs Laboratories, ($480$, $520$)).

\subsection{Sample preparation}
Stock cultures of Hela cells were maintained in RPMI (Roswell Park Memorial Institute medium)  supplemented with $10\%$ (v/v) deactivated  fetal bovine serum, $2\,$mM L-glutamine and $50\,\mu$g/ml gentamycin (complete medium, Lonza) at $37^{\circ}$C in a humidified atmosphere of $95\%$ air/$5\%$ $\textrm{CO}_2$.
Cells were recovered by trypsinisation (trypsin-EDTA, Lonza).

Fluorescent microspheres (Bangs Laboratories) were used, with a mean diameter of $0.51\,\mu$m (FC03F). The suspensions of beads were homogenized (rotation) overnight at $4^{\circ}$C before use.
$5 \times 10^5$ cells were mixed with $10\,\mu$l of  bead suspension and $4\,$ml complete culture medium, plated  in $10\,\mbox{cm}^2$ culture wells and incubated for $24$ hours  at $37^{\circ}$C in a humidified atmosphere of $95\%$ air/$5\%$ $\textrm{CO}_2$. At the end of the incubation the cells were recovered by trypsinisation. To eliminate the beads not adsorbed by the cells, cells were diluted in $2\,$ml medium, put in a $3\,\mu$m pore size cell culture insert (Greiner Bio-one) and briefly centrifuged at $200\,$g. Cells were recovered by washing the insert with culture medium, plated on glass coverslips at a density of  $5 \times 10^4$ cells$/\mbox{cm}^2$, and incubated overnight at $37^{\circ}$C under an atmosphere of $5\%$ $\textrm{CO}_2$. The coverslips were then rinsed $3$ times with phosphate buffered saline (PBS), fixed with $4\%$ Paraformaldehyde (PFA) during $10$ minutes and rinsed again with PBS.

\subsection{Acquisition parameters}
All imaging experiments were performed at a modulation frequency $\nu_m$ of $17\,$kHz, and the first, second ($34\,$kHz) and third ($51\,$kHz) harmonics were demodulated. The images containing only the first and second harmonics were obtained with $10\,$mW excitation power on the sample. To obtain the images of the third harmonic as well as the linear combination, the excitation power was increased to $25\,$mW to produce a sufficiently strong signal. The complete 2D field of view $(x,y)$ of $508.68 \times 508.68\,\mu\mbox{m}^2$ is sampled at $4096 \times 4096$ pixels with a pixel dwell time of $200\,\mu$s, which corresponds to $3.4$ modulation periods at $17\,$kHz, $6.8$ at $34$\,kHz and $10.2$ at $51\,$kHz. $(x,y,z)$ scans are stacks where each \mbox{$(x,y)$-slice} (same field of view as for 2D images) is sampled at $2048\times 2048$ pixels.

\section{Results}

\subsection{Microspheres on a coverslip}
\begin{figure}[h]
\centering
\includegraphics[width=0.5\textwidth]{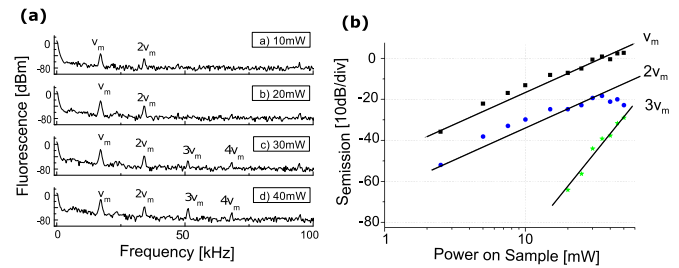}
\caption{Harmonics in the detected fluorescence signal from $0.19\mu$m microspheres on a coverslip. (a) Frequency spectra of the fluorescence signal emitted by the microspheres for increasing average power. (b) Spectral intensities of the three first harmonics in the fluorescence signal. Linear fits of the data are shown as black lines. The particle concentration was $10\,$mg/mL and the frequency of the excitation modulation $\nu_m$ was $17\,$kHz.}
\label{Figure3}
\end{figure}

To demonstrate the resolution improvement of our method, $0.19\,\mu$m microspheres were imaged. First we determined the power needed to obtain saturation by measuring the harmonics in the fluorescence signal. To this end, we analyzed the signal from the PMT with an electronic spectrum analyzer (ESA) ($8564$E, Hewlett Packard), and increased the illumination power until the harmonics rose above the noise floor of the ESA as shown in Fig.~\ref{Figure3}(a). Note that, due to the narrowband detection of the lock-in amplifier, the actual noise floor of the imaging system is expected to be lower than the noise floor measured by the ESA ($-70\,$dBm).

The third harmonic clearly emerged from the noise for an illumination power of approximately $30\,$mW on the sample. At this illumination power, the signal-to-noise ratio was measured to be $15\,$dB and was sufficient to acquire an image. The demodulated fluorescence intensity at a frequency of $\nu_m$ ($17\,$kHz), $2\nu_m$ ($34\,$kHz) and $3\nu_m$ ($51\,$kHz) is shown in Fig.~\ref{Figure3}(b). As expected from Eq.~(\ref{harmonicsSups}) and Fig.~\ref{Figure1}(b), the curves of the fundamental and second harmonics are linear and have the same slope, while the slope of the third harmonic curve is twice as steep.

\begin{figure}[h]
\centering
\includegraphics[width=0.4\textwidth]{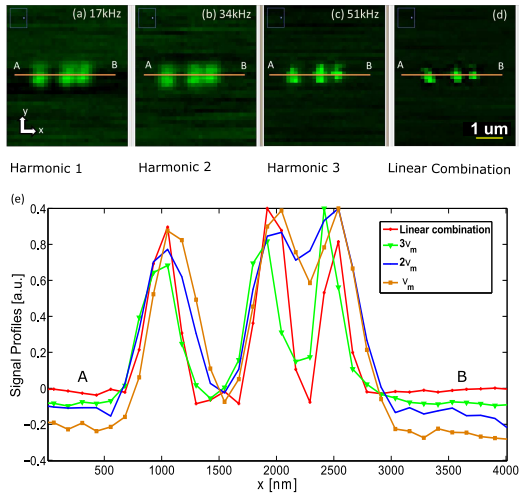}
\caption{Characterization of the lateral resolution improvement using $0.19\,\mu$m microspheres on a coverslip. (a-d) Images in the $(x,y)$ plane reconstructed with the first, second, and third harmonics and the linear combination of these three harmonics. The \mbox{$(x,y)$-scan} size is $48.36 \times 56.54\,\mu\mbox{m}^2$ ($390\times456\,$ pixels) resulting in a total acquisition time of $\sim35\,$s. (e) Signal profiles along the orange line in (a-d). The background intensity was measured in an empty area and subtracted everywhere.}
\label{Figure4}
\end{figure}

For the fundamental ($17\,$kHz) and the second harmonic ($34\,$kHz) [Fig.~\ref{Figure4}(a) and (b)], the spatial resolution in $x$ and $y$ was measured to be the same as in conventional MPM. The images obtained using the third harmonic [Fig.~\ref{Figure4}(c)] and the linear combination of the three first harmonics given by Eq.~(\ref{CombiLi}) [Fig.~\ref{Figure4}(d)] clearly show a resolution improvement. The intensity profiles shown in Fig.~\ref{Figure4}(e) confirm this improvement: for the first and the second harmonics, the two microspheres are not resolved, they can just be resolved by the third harmonic. The linear combination resolves the two microspheres even better.

The full-width at half-maximum (FWHM) was measured on the leftmost microsphere to be approximately $470\,$nm for the two first harmonics, $340\,$nm for the third and $285\,$nm for the linear combination signal. These measurements fit with the estimated theoretical improvement of the resolution.

\begin{figure}[h]
\centering
\includegraphics[width=0.35\textwidth]{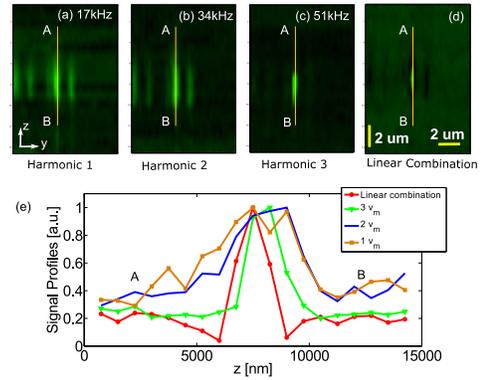}
\caption{Characterization of the axial resolution improvement using $0.19\,\mu$m microspheres on a coverslip. (a-d) Images in the $(y,z)$ plane reconstructed with the first, second, and third harmonics and the linear combination of these three harmonics. The \mbox{$(y,z)$-sections} are taken from $(x,y,z)$-stacks consisting of 19 $(x,y)$-slices of $49.10 \times 58.53\,\mu\mbox{m}^2$ ($192\times236\,$ pixels), resulting in a total acquisition time of 3 minutes per 3D stack. (e) Signal profiles along the yellow line in (a-d). The background intensity was measured in an empty area and subtracted everywhere.}
\label{Figure5}
\end{figure}

The same analysis was performed in the axial direction to demonstrate the improvement of the axial resolution. Figure~\ref{Figure5} shows the four different $(x,z)$ planes obtained with the first, second, and third harmonics and the linear combination. The results confirm that the axial resolution is also improved using saturation of the excitation. The axial FWHM was measured to be approximately $3.4\,\mu\mbox{m}$ for the two first harmonics, $2.1\,\mu\mbox{m}$ for the third and $1.8\,\mu\mbox{m}$ for the linear combination signal. Again, these values match the estimated theoretical resolution improvement.

\subsection{Microspheres in HeLa cells}
\begin{figure}[h]
\centering
\includegraphics[width=0.5\textwidth]{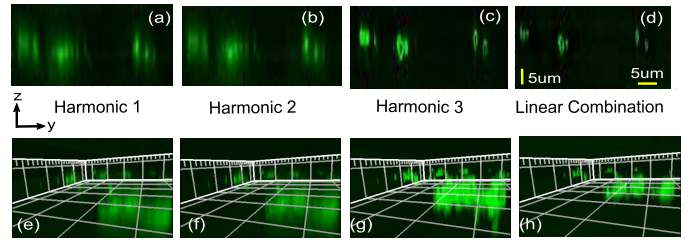}
\caption{Microspheres of $0.51\,\mu$m in HeLa cells. (a-d) Images in the $(y,z)$ plane show a clear improvement of the resolution in the lateral and axial dimensions. (e-h) 3D renderings for the first, second and third harmonics and the linear combination of the three first harmonics. The $(x,y,z)$-stacks consisted of 17 $(x,y)$-slices of $169.6 \times 169.6\,\mu\mbox{m}^2$ ($800\times800\,$ pixels), resulting in a total acquisition time of 36 minutes per 3D stack.}
\label{Figure6}
\end{figure}

To demonstrate the applicability of this multiphoton super-resolution technique to biological samples, we went on to image $0.51\,\mu\mbox{m}$ microspheres incorporated in HeLa cells. Figure \ref{Figure6} shows $(y,z)$ sections of 3D images constructed from the first, second, and third harmonics and the linear combination of those three harmonics. 3D renderings were generated with the Imaris software (Bitplane AG).

As for the microspheres on a coverslip, an improvement of the resolution in all three dimensions is observed when imaging at the third harmonic and a further improvement is obtained by the linear combination of the three first harmonics.


\section{Discussion and conclusions}
In this work, we demonstrated experimentally that saturation of fluorescence excitation can be exploited to obtain a 3D resolution improvement in MPM. Implementation of this technique on an existing MPM system is straightforward since it only requires modulating the excitation beam and demodulating the detected fluorescence signal. The method is also cost-effective because it does not require additional laser sources or microscope objectives to improve the 3D resolution. Moreover, the technique does not employ delicate 3D beam-shaping optics, which are sensitive to misalignment over time.

Images of fluorescent microspheres on a coverslip and integrated in HeLa cells show that resolution is indeed improved in all three spatial dimensions by a factor of $\sqrt{2}$ when imaging at the demodulated third harmonic of the detected fluorescence signal. Furthermore, we showed that the 3D resolution can be improved by an additional factor of $\sqrt{3/2}$ when the appropriate linear combination is made of the first three harmonics. This linear combination of harmonics is made by an analog circuit such that no subsequent signal processing is required. For comparison, our super-resolution MPM obtains the same lateral resolution as a $400\,$nm illuminated linear confocal microscope (assuming the same beam characteristics and objective), while conserving the advantages of classical MPM (i.e. near-infrared illumination and increased depth of penetration).

To obtain the required saturation regime and get a sufficiently high signal-to-noise ratio for the harmonic signals, an average illumination power of $25\,$mW at $880\,$nm was delivered on the sample. Interestingly, this rather modest illumination power suffices to obtain a near twofold 3D resolution enhancement in MPM. Moreover, since the sample is illuminated only at near-infrared wavelengths, our approach is likely to be less invasive, in terms of photo-toxicity, than the super-resolution techniques based on STED.

With the present acquisition parameters, relatively few modulation periods are measured at every image voxel. Even though the resulting signal-to-noise ratio was sufficient for imaging, there is room for improvement by increasing the number of periods of modulation of the excitation signal per pixel. This can be done by increasing the pixel dwell time (if the acquisition time is not critical) or by using a higher modulation frequency (but keeping a sufficient number of pulses per modulation period). An improved signal-to-noise ratio could allow using a lower illumination power while still obtaining a sufficiently strong signal to reconstruct the image. This could be particularly interesting to investigate for future work, especially for \textit{in vivo} applications where the power damage threshold is critical.

Theoretically, the resolution improvement is unlimited since higher harmonics than the third one can be used. However, as for all super-resolution methods based on saturation of the excitation or STED, the attainable resolution is ultimately limited by the damage threshold of the sample. At high illumination powers, photobleaching and phototoxicity become critical in biological samples. This problem can be reduced using fluorophores with high two-photon absorption cross section or quantum dots, provided that their fluorescence lifetime is sufficiently short compared to the modulation period.


\section*{Acknowledgments}
This work was supported by the Institut Interuniversitaire des Sciences Nucléaires (IISN), Grant No 4.4510.09 and by the Belgian Science Policy Office (BELSPO) Interuniversity Attraction Pole (IAP), Project IAP 7/35 Photonics@be.

Fr\'ed\'erique Vanholsbeeck acknowledges the support of a University of Auckland grant for research and study leave, a ``Brains (Back) to Brussels" grant (B2B 2012-2-08) from the Brussels institute for research and innovation (Innoviris) and a ``Missions Scientifiques Professeurs Etrangers" grant from F.N.R.S. that supported her stay at the Universite libre de Bruxelles.



\end{document}